\magnification=\magstep1
\overfullrule=0pt
\parskip=6pt
\baselineskip=15pt
\headline={\ifnum\pageno>1 \hss \number\pageno\ \hss \else\hfill \fi}
\pageno=1
\nopagenumbers
\hbadness=1000000
\vbadness=1000000

\input epsf

\vskip 25mm
\vskip 25mm
\vskip 25mm

\centerline{\bf PERMUTATION WEIGHTS AND MODULAR POINCARE
POLYNOMIALS}
\centerline{\bf FOR AFFINE LIE ALGEBRAS}

\vskip 5mm

\centerline{\bf Meltem Gungormez} \centerline{Dept. Physics, Fac.
Science, Istanbul Tech. Univ.} \centerline{34469, Maslak, Istanbul,
Turkey } \centerline{e-mail: gungorm@itu.edu.tr}

\vskip 3mm

\centerline{\bf Hasan R. Karadayi} \centerline{Dept. Physics, Fac.
Science, Istanbul Tech. Univ.} \centerline{34469, Maslak, Istanbul,
Turkey } \centerline{e-mail: karadayi@itu.edu.tr} \centerline{and}
\centerline{Dept. Physics, Fac. Science, Istanbul Kultur University}
\centerline{34156, Bakirkoy, Istanbul, Turkey }

\vskip 3mm

\centerline{\bf{Abstract}}

Poincare Polynomial of a Kac-Moody Lie algebra can be obtained by
classifying the Weyl orbit $W(\rho)$ of its Weyl vector $\rho$. A
remarkable fact for Affine Lie algebras is that the number of
elements of $W(\rho)$ is finite at each and every depth level though
totally it has infinite number of elements. This allows us to look
at $W(\rho)$ as a manifold graded by depths of its elements and
hence a new kind of Poincare Polynomial is defined. We give these
polynomials for all Affine Kac-Moody Lie algebras, non-twisted or
twisted. The remarkable fact is however that, on the contrary to the
ones which are classically defined,these new kind of Poincare
polynomials have modular properties, namely they all are expressed
in the form of eta-quotients. When one recalls Weyl-Kac character
formula for irreducible characters, it is natural to think that this
modularity properties could be directly related with Kac-Peterson
theorem which says affine characters have modular properties.

Another point to emphasize is the relation between these modular
Poincare Polynomials and the Permutation Weights which we previously
introduced for Finite and also Affine Lie algebras. By the aid of
permutation weights, we have shown that Weyl orbits of an Affine Lie
algebra are decomposed in the form of direct sum of Weyl orbits of
its horizontal Lie algebra and this new kind of Poincare Polynomials
count exactly these permutation weights at each and every level of
weight depths.

\eject

\vskip 3mm
\noindent {\bf{I.\ INTRODUCTION }}
\vskip 3mm

We know that any affine Lie algebra $\widehat{G}_N$ is related with a
finite Lie algebra $G_N$ which is called its horizontal Lie algebra.
Let $\lambda_i$'s and $\alpha_i$'s be respectively the fundamental
weigths and simple roots of horizontal Lie algebra $G_N$ where
$i=1,2 \dots ,N$. They are determined by
$$ {2 \ \kappa(\lambda_i,\alpha_j) \over \kappa(\alpha_i,\alpha_j) }
\equiv \delta_{i,j}  $$ where $\kappa(,)$ is symmetric scalar
product which is known always to be exist via the relation
$$ {2 \ \kappa(\alpha_i,\alpha_j) \over \kappa(\alpha_i,\alpha_j) } \equiv (A_N)_{i,j} $$
where $A_N$ is the Cartan matrix of $G_N$. We follow the book of Humphreys {\bf [1]}
for finite and Kac {\bf[2]} for Kac-Moody Lie algebras.

Let $\widehat{A}_N$ be the Cartan matrix and $\alpha_0$ the extra simple root of $\widehat{A}_N$.
Its dual $\lambda_0$ is to be introduced by hand
via the relations
$$ \eqalign{
&\kappa(\lambda_0,\lambda_0) = 0  \cr
&\kappa(\lambda_0,\alpha_0) = 1  } $$
due to the fact that $\widehat{A}_N$ is singular. Note also that
$$ \kappa(\lambda_0,\alpha_i) = 0 \ \ , \ \ i=1,2, \dots ,N.  $$
and hence the name {\bf horizontal} for $G_N$. Affine Lie algebras are also characterized
by the existence of a unique isotropic root $\delta$ defined by
$$ \delta = \sum_{\mu=0}^N \ k_\mu \ \alpha_\mu  $$
where $k_\mu$'s are known to be {\bf Kac labels} of $\widehat{G}_N$. Let $ W_{\widehat{G}_N}$
be the weight lattice of $\widehat{G}_N$. For any element $\widehat{\lambda} \in \widehat{G}_N$,
we know the following decomposition is always valid:
$$ \widehat{\lambda} = \lambda + k \ \lambda_0 - M \ \delta  \eqno(I.1)$$
where the {\bf level} \ k \ is constant for the Weyl orbit $W(\widehat{\lambda})$
and the $\bf{depth}$ \ M \ is always defined to take values
$$ M=0,1, \dots , \infty . $$
For any fixed value of M, let us now define $ W_M(\widehat{\lambda})$ to be the set of weights
with the form (I.1). It is known that the orders of these sets are always finite, that is
$$ \mid W_M(\widehat{\lambda}) \mid  <  \infty      \eqno(I.2) $$
though their completion and hence the order of $
W(\widehat{\lambda})$ is infinite. In view of (I.2), we suggest that
$ W(\widehat{\lambda})$ can be considered as a manifold graded by
weight depths M and hence a Poincare polynomial $ Q(\widehat{G}_N)$
is attributed by the following definition:
$$ Q(\widehat{G}_N) \equiv \sum_{M=0}^\infty \mid W_M(\widehat{\lambda}) \mid
\ t^M  \eqno(I.3) $$ where t is taken to be an indeterminate here
and also in the following. It is a priori clear that these
polynomials are quite different from Affine Poincare polynomials $
P(\widehat{G}_N)$ which are known to be defined by
$$ P(\widehat{G}_N)=P(G_N) \prod^N_{i=1} {1 \over 1-t^{d_i-1}} \ .  \eqno(I.4) $$
by Bott theorem {\bf [3]}. In (I.4), $ P(G_N) $ is the Poincare
polynomial and $d_i$'s are exponents of $G_N$.

As we have shown in another work{\bf [4]}, an explicit calculation
of Poincare polynomials of Hyperbolic Lie algebras can be carried
out by classifying the Weyl orbit $W(\rho)$ in terms of
lengths{\bf[5]} of Weyl group elements. Such a calculation is
extended in a direct way to a classification in terms of weight
depths M. For simply-laced affine Lie algebras,

$$ \widehat{G}_N = A_N^{(1)}, D_N^{(1)} , E_6^{(1)} , E_7^{(1)} , E_8^{(1)} $$

\noindent depicted in p.54 of [2], these calculations give the
result

$$ Q(\widehat{G}_N)= { \mid W_{G_N} \mid  \over P_N \ R_N  }  \eqno(I.5)  $$

\noindent where

$$ \eqalign{
&P_N = \prod_{i=1}^N  \prod_{k=0}^\infty (1-q^{h k + d_i})  \cr
&R_N =\prod_{k=0}^\infty \prod_{s=0}^\infty (1+q^{ 2^s (2 k + 1) h^\vee})^{(s+1) N} }  \eqno(I.6) $$

\noindent In above expressions, \ h \ and $h^\vee$ are coxeter and co-coxeter numbers of $G_N$ and
$\mid W_{G_N} \mid$ is the order of Weyl group $  W_{G_N} $ of finite Lie algebra $G_N$.

Although similar expressions could be obtained for a complete list of affine Lie algebras,
this will be presented in the next section in which we expose modular properties of
Q-Poincare polynomials defined in (I.3).

\vskip 3mm
\noindent {\bf{II.\ POINCARE POLYNOMIALS AS
ETA-QUOTIENTS} }
\vskip 3mm

There is quite vast litterature {\bf [6]} on eta-quotients which are rational products of Dedekind
eta functions with several arguments. Their relation with finite groups is also studied {\bf [7]}. Let
$$ \varphi(q)=\prod_{i=1}^\infty (1-q^i) $$
be Euler product and
$$ \eta(\tau) \equiv q^{1/24} \ \varphi(q) $$
Dedekind eta function where $ q=e^{2 \pi i \tau} $. An eta-quotient is defined {\bf [8]} to be a function
$ f(\tau)$ of the form
$$ f(\tau) \equiv \prod_{i=1}^d \eta(s_i \tau)^{r_i}  \eqno(II.1)$$
where $\{ s_1,s_2, \dots ,s_d \}$ is a finite set of positive integers and $r_1,r_2, \dots ,r_d $ are arbitrary
integers. Let us denote the collection of integers $ r_1, s_1, r_2, s_2, \dots ,r_d, s_d $ defining $f(\tau)$
by the formal product
$$ g=s_1^{r_1} s_2^{r_2} \dots s_d^{r_d} \eqno(II.2) $$
and write $\eta_g(\tau)$ for the corresponding eta-quotient (II.1).

What's important here is to emphasize eta-products are in general
meromorphic modular forms of weight $k \equiv { 1 \over 2}
\sum_{i=1}^d r_i $ and multiplier system for some congruence
subgroup of $SL_2(Z)$ . This study is however outside the scope of
this paper so we will only give here the complete list of Poincare
series defined above in the notation of (II.2). To this end, we
define
$$ Q(\widehat{G}_N)= \mid W_{G_N}\mid \
q^{{1 \over 24} \ \phi(\widehat{G}_N)} \ \eta_{g(\widehat{G}_N)}  .
\eqno(II.3)
$$ The phase factors $ q^{{1 \over 24} \ \phi(\widehat{G}_N)} $ which stem
from the difference between definitions of Euler product and
$\eta$-function will also be given. Our results are given in the
following Table-1 for non-twisted types in Kac's table Aff 1 ({\bf
p.54 of [2]}) and in Table-2 for twisted types of Table Aff 2 ({\bf
p.55 of [2]}):

\eject

\centerline{ Table-1}

$$ \eqalign{
&g(A_N^{(1)})=(h^\vee)^{(N+1)} \ 1^{-1} \ \ \ \ \ \ \ \ \ \ \ \ \ \
\ \ \ \ \ \ \ \ \ \ \ , \ \ \phi(A_N^{(1)})=-(N+1) h^\vee + 1 \cr
&g(B_N^{(1)})=(2 h^\vee)^1 \ (h^\vee)^{(N-1)} \ 2^1 \ 1^{-1}  \ \ \
\ \ \ \ \ \ \ \ , \ \ \phi(B_N^{(1)})=-(N+1) h^\vee - 1 \cr
&g(C_N^{(1)})=(2 h^\vee)^{(N-1)} \ (h^\vee)^1 \ 2^1 \ 1^{-1}  \ \ \
\ \ \ \ \ \ \ \ , \ \ \phi(C_N^{(1)})=-(N+1) h^\vee - 1 \cr
&g(D_N^{(1)})=(h^\vee)^{(N+1)} \ ({1 \over 2} h^\vee)^{-1} \ 2^1 \
1^{-1}  \ \ \ \ \ \ \ \  , \ \ \phi(D_N^{(1)})=-(N+{1 \over 2})
h^\vee -1 \cr &g(G_2^{(1)})=12^1 \ 6^{-1} \ 4^1 \ 3^1 \ 2^1 \ 1^{-1}
\ \ \ \ \ \ \ \ \ \ \ \ \ \ \ \  , \ \ \phi(G_2^{(1)})=-(2+1) \ 6 \
\  + (6-2)      \cr &g(F_4^{(1)})=(18)^2 \ 9^2 \ 6^{-1} \ 3^1 \ 2^1
\ 1^{-1} \ \ \ \ \ \ \ \ \ \ \ \ \ \ , \ \ \phi(F_4^{(1)})=-(4+1) 12
\ + (12-4)      \cr &g(E_6^{(1)})=12^7 \ 6^{-1} \ 4^{-1} \ 3^1 \ 2^1
\ 1^{-1} \ \ \ \ \ \ \ \ \ \ \ \ \ \  , \ \ \phi(E_6^{(1)})=-(6+1)
12 \ + (12-6)    \cr &g(E_7^{(1)})=18^8 \ 9^{-1} \ 6^{-1} \ 3^1 \
2^1 \ 1^{-1} \ \ \ \ \ \ \ \ \ \ \ \ \ \ , \ \
\phi(E_7^{(1)})=-(7+1) 18 \ + (18-7)     \cr &g(E_8^{(1)})=30^9 \
15^{-1} \ 10^{-1} \ 6^{-1} \ 5^1 \ 3^1 \ 2^1 \ 1^{-1} \ \ , \ \
\phi(E_8^{(1)})=-(8+1) 30 \ + (30-8)    \cr } $$

\vskip  5mm

\centerline{ Table-2}
$$ \eqalign{
g(A_2^{(2)})&=12^1 \ \ 6^{-1} \ 4^{-1} \ 3^1 \ 2^2 \ 1^{-1} \ \ \ \
\ \ \ \ \ \ \ \ \ \ \ \ \ \ \ \ \ , \ \ \phi(A_N^{(2)})= -8 \cr
g(A_{2 N}^{(2)})&=(4 h^\vee)^1 \ (2 h^\vee)^{(N-2)} \ (h^\vee)^1 \
4^{-1} \ 2^2 \ 1^{-1} \ \ \  , \ \ \phi(A_{2 N}^{(2)})=-(2 N+1)
h^\vee + 1 \cr g(A_{2 N-1}^{(2)})&=(2 h^\vee)^2 \ (h^\vee)^{(N-3)} \
N^1 \ 2^1 \ 1^{-1} \ \ \ \ \ \ \ \ \ \ \ \ \ , \ \ \phi(A_{2
N-1}^{(2)})=-(N+1) h^\vee - (N + 1) \cr g(D_{N+1}^{(2)})&=(2
h^\vee)^N \ 4^{-1} \ 2^2 \ 1^{-1} \ \ \ \ \ \ \ \ \ \  \ \ \ \ \ \ \
\ \ \ \ \ \ \ \ \ , \ \ \phi(D_{N+1}^{(2)})=-2 N h^\vee + 1 \cr
g(E_6^{(2)})&=12^1 \ 8^{-1} \ 6^{-1} \ 4^1 \ 3^1 \ 2^1 \ 1^{-1} \ \
\ \ \ \ \ \ \ \ \ \ \ \ \ \ \ \ , \ \ \phi(E_6^{(2)})=-6  \cr
g(D_4^{(3)})&=18^2 \ 9^{-1} \ 6^{-1} \ 3^2 \ 2^1 \ 1^{-1} \ \ \ \ \
\ \ \ \ \ \ \ \ \ \ \ \ \ \ \ \ \ , \ \ \phi(D_4^{(3)})=-28  \cr }
$$

\vskip  2mm

At first glance, the examples $G_2^{(1)} , F_4^{(1)} , E_6^{(1)} ,
E_7^{(1)} , E_8^{(1)} $ and $ D_4^3$ are interesting due to a
theorem {\bf [6]} concerning modular forms for some congruence
subgroups of $SL_2(Z)$ . We leave however such a study in a
subsequent paper.

\eject

\vskip 6mm

\noindent {\bf{III.\ POINCARE POLYNOMIALS AND PERMUTATION WEIGHTS FOR
AFFINE LIE ALGEBRAS} }

\vskip 3mm

We have defined Permutation Weights previously for finite Lie
algebras {\bf [9]} and also Affine Lie algebras {\bf [10]}. In these
works, it is shown that permutation weights can be calculated
explicitly by the aid of a contructive corollary ({\bf p.7 of
[10]}).

Here, it is shown that the polynomials
$$ { Q(\widehat{G}_N) \over \mid W_{G_N}\mid  } \eqno(III.1)   $$
count permutation weights at each and every depth level, as will be
exemplified in what follows. The present method provides a direct
way to find permutation weights by explicit calculation of affine
Weyl group elements to which the permutation weights are obtained.
One could say that this is not generally so practical since it needs
explicit calculations of Weyl group elements. The present method is
however presented here as an independent investigation of the
previous one.

All our Lie algebraic definitions are as in the sec.I. Let us first
briefly remember our previous definition and determination {\bf
[10]} of permutation weights. Let $\widehat{\Lambda}^+$ and
$\lambda^+$ be dominant weights of $\widehat{G}_N$ and $G_N$
respectively, $W(\widehat{\Lambda}^+)$ and $W(\lambda^+)$  be
corresponding Weyl orbits. We know that all the elements of $
W(\widehat{\Lambda}^+) $ has the form (I.1) and among them the
permutation weights are defined by the following specific form:
$$  \lambda^+ + k \ \lambda_0 - M \ \delta \ \ , \ \ M=1,2, \dots  . \eqno(III.2) $$
In (III.2), for each and every value of M, we define ${\cal
P}_M(\widehat{\Lambda}^+) $ to be the set of permutation weights of
$\widehat{\Lambda}^+ $ and $ \mid {\cal P}_M(\widehat{\Lambda}^+)
\mid $ be its order. Let also note that (III.1) can always be
expressed in the following form:

$$ { Q(\widehat{G}_N) \over \mid W_{G_N}\mid  } =
\sum_{M=0}^\infty \ c_M \ q^M  \eqno(III.3) $$
where $c_M$'s are positive integers, $c_0=1$ and q is an indeterminate. One can show that

$$ \mid {\cal P}_M(\widehat{\Lambda}^+) \mid \ = c_M  \ \ , \ \ M=1,2, \dots
\eqno(III.4)  $$
And hence $Q(\widehat{G}_N)$ states that all the elements $ \lambda + k \ \lambda_0 - M \ \delta $
are belong to $ W(\widehat{\Lambda}^+) $ where $ \lambda \in W(\lambda^+) $. In other words, if

$$ \lambda^+ + k \ \lambda_0 - M \ \delta \in {\cal P}_M(\widehat{\Lambda}^+)   $$
is exist for any M, then one finds that

$$ W(\lambda^+) + k \ \lambda_0 - M \ \delta \in W(\widehat{\Lambda}^+)   $$
due to existence of Poincare series given in Table-1 and also Table-2. The existence of
these new kind of Poincare series is the existence of permutation weights. One can formally
say that this gives us an explicit way to decompose any Weyl orbit of an Affine Lie algebra as
a direct sum of Weyl orbits of its horizontal Lie algebra. This reflects our main point of view
to introduce permutation weights.

It is now useful to proceed in an example for which all our general framework is to be reflected.
Let us consider the simply laced, affine Kac-Moody Lie algebra $E_6^{(1)}$ with the following
Dynkin diagram:

\vskip 6mm

\epsfxsize=5cm \centerline{\epsfbox{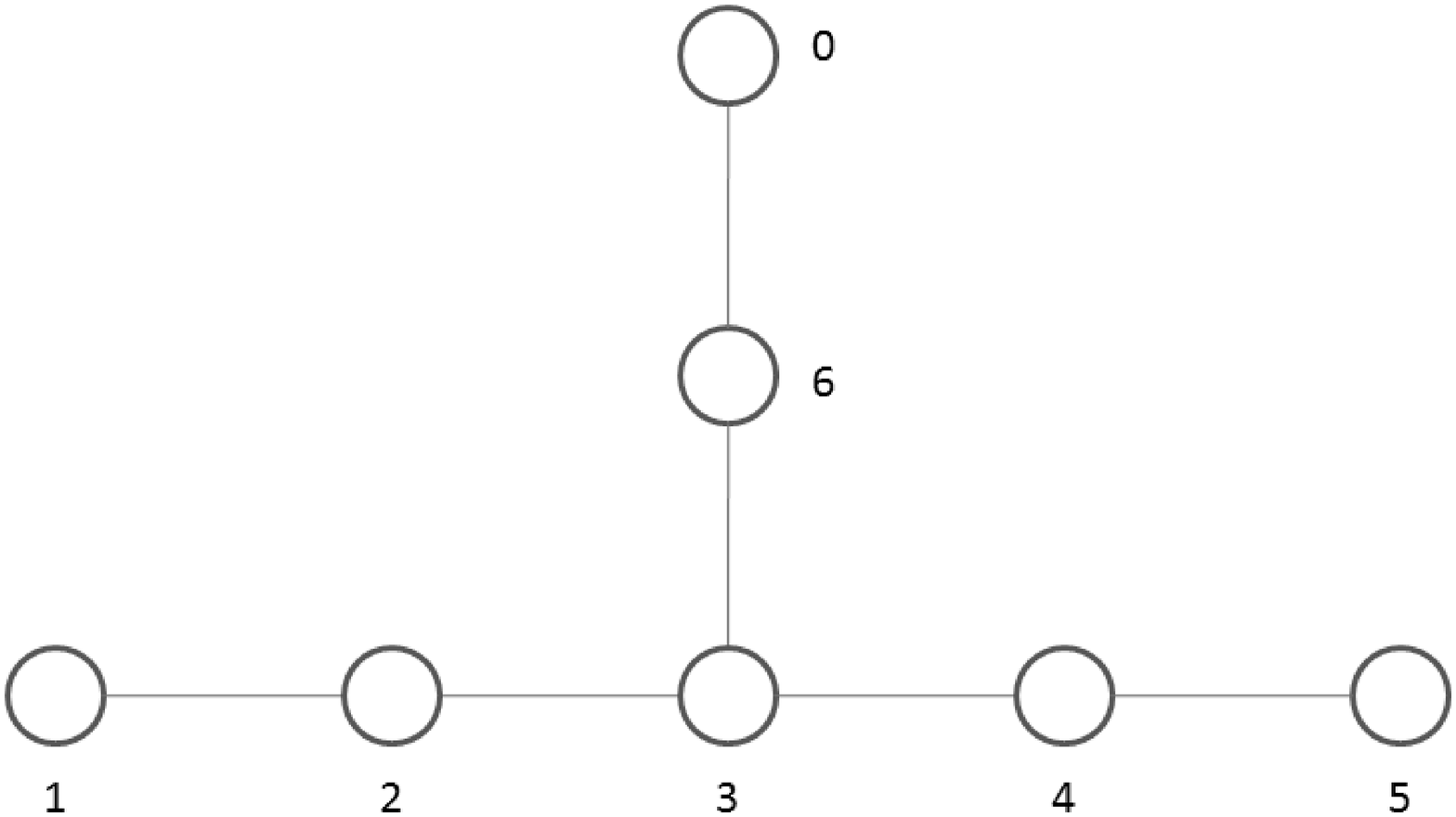}}

\vskip 3mm

From Table-1 of Sec.II, one finds that the similar of (III.3) is
$$ { Q(E_6^{(1)}) \over \mid W_{E_6} \mid }  =
1 + q + q^2 + q^3 + 2 \ q^4 + 3 \ q^5 + 3 \ q^6 + 4 \ q^7 + 6 \ q^8 + 7 \ q^9 + \dots  \eqno(III.1) $$
In view of (III.3) and (III.4), the following Table-3 is trivial:

\vskip 3mm

\centerline { Table-3}

$$ \vbox{\tabskip=0pt \offinterlineskip
\halign to250pt{\strut#& \vrule#\tabskip=0em plus2em& \hfil#&
\vrule#& \hfil#& \vrule# \tabskip=0pt\cr \noalign{\hrule} & &
\omit\hidewidth M \hidewidth&
  & \omit\hidewidth \ \ \ 1 \ \  2 \ \  3 \ \ 4 \ \ 5 \ \ 6 \ \ 7 \ \ 8 \ \ 9  \hidewidth& \cr \noalign{\hrule}
& & $ c_M$ &  & \  \ 1 \ \  1 \ \  1 \ \ 2 \ \ 3 \ \ 3 \ \ 4 \ \ 6 \ \
7& \cr \noalign{\hrule} \hfil\cr}}
$$


\eject

Let $W(E_6^{(1)})$ be the Weyl group of $E_6^{(1)}$. Then, all the elements which give us the permutation
weights numbered in above Table-3 are given explicitly as in the following:
$$ \eqalign{
\Sigma_{1,1}&=\sigma_{0}  \cr \cr \Sigma_{2,1}&=\sigma_{0,6} \cr \cr
\Sigma_{3,1}&=\sigma_{0,6,3} \cr \cr \Sigma_{4,1}&=\sigma_{0,6,3,2}
\ \  , \ \ \Sigma_{4,2}=\sigma_{0,6,3,4} \cr \cr
\Sigma_{5,1}&=\sigma_{0,6,3,2,1} \ \ , \ \
\Sigma_{5,2}=\sigma_{0,6,3,2,4} \ \ , \ \
\Sigma_{5,3}=\sigma_{0,6,3,4,5} \cr \cr
\Sigma_{6,1}&=\sigma_{0,6,3,2,1,4} \ \ , \ \
\Sigma_{6,2}=\sigma_{0,6,3,2,4,3} \ \ , \ \
\Sigma_{6,3}=\sigma_{0,6,3,2,4,5}  \cr \cr
\Sigma_{7,1}&=\sigma_{0,6,3,2,1,4,3} \ \ , \ \
\Sigma_{7,2}=\sigma_{0,6,3,2,1,4,5}  \cr
\Sigma_{7,3}&=\sigma_{0,6,3,2,4,3,5} \ \ , \ \
\Sigma_{7,4}=\sigma_{0,6,3,2,4,3,6}   \cr \cr
\Sigma_{8,1}&=\sigma_{0,6,3,2,1,4,3,2} \ \ , \ \
\Sigma_{8,2}=\sigma_{0,6,3,2,1,4,3,5}   \cr
\Sigma_{8,3}&=\sigma_{0,6,3,2,1,4,3,6} \ \ , \ \
\Sigma_{8,4}=\sigma_{0,6,3,2,4,3,5,4}   \cr
\Sigma_{8,5}&=\sigma_{0,6,3,2,4,3,5,6} \ \ , \ \
\Sigma_{8,6}=\sigma_{0,6,3,2,4,3,6,0}    \cr \cr
\Sigma_{9,1}&=\sigma_{0,6,3,2,1,4,3,2,5} \ \ , \ \
\Sigma_{9,2}=\sigma_{0,6,3,2,1,4,3,2,6}  ,  \cr
\Sigma_{9,3}&=\sigma_{0,6,3,2,1,4,3,5,4} \ \ , \ \
\Sigma_{9,4}=\sigma_{0,6,3,2,1,4,3,5,6}  ,  \cr
\Sigma_{9,5}&=\sigma_{0,6,3,2,1,4,3,6,0} \ \ , \ \
\Sigma_{9,6}=\sigma_{0,6,3,2,4,3,5,4,6}  ,  \cr
\Sigma_{9,7}&=\sigma_{0,6,3,2,4,3,5,6,0}   } $$ We assume here that
Weyl group elements are expressed in the form of \break
$\sigma_{\mu_1,\mu_2, \dots \mu_k} \equiv \sigma_{\mu_1} \
\sigma_{\mu_2} \dots \sigma_{\mu_k} $ where $\sigma_\mu$'s are
simple Weyl reflections which are defined to be the Weyl group
elements with respect to simple roots $\alpha_\mu$ of $E_6^{(1)}$
with $\mu=0,1,2, \dots 6$. On the left-hand side, actions of Weyl
group elements are defined on $E_6^{(1)}$ weight lattice by
$$ \Sigma_{M,c_M}(\widehat{\Lambda}^+) \equiv \widehat{\Lambda}^+ +
\kappa(\widehat{\Lambda}^+,\delta) \ \lambda_0 - M \ \delta  . $$

\vskip3mm
\noindent{\bf {REFERENCES}}
\vskip3mm

\item [1] J. E. Humphreys, Introduction to Lie Algebras and Representation Theory,
\item \ \ \ \ Springer-Verlag, 1972
\item [2] V. Kac, Infinite Dimensional Lie Algebras, Cambridge University Press, 1982
\item [3] R. Bott, An Application of the Morse Theory to the topology of Lie-groups,
\item \ \ \ \ \ Bull. Soc. Math. France 84 (1956) 251-281
\item [4] M.Gungormez  and H.R.Karadayi, On Poincare Polynomials of Hyperbolic Lie
\item \ \ \ \ \ Algebras, arXiv:0706.2563
\item [5] J. E. Humphreys, Reflection Groups and Coxeter Groups,
\item \ \ \ \ \ Cambridge University Press, 1990
\item [6] Y. Martin, Multiplicative Eta-Quotients. Trans.Amer.Math.Soc.348,
\item \ \ \ \ \ no.12 (1996), 4825-4856.
\item \ \ \ Y.Martin, K.Ono,Eta-Quotients and Elliptic Curves, Proc.Am.Math.Soc.124,
\item \ \ \ \ \ no:11(1997), 3169-3176
\item \ \ \  M. Newman, Construction and application of a class of modular functions,
\item \ \ \ \ \ J.London Math.Soc.(3)7 (1957), 334-350.
\item \ \ \  M. Newman, Construction and application of a class of modular
functions (II),
\item \ \ \ \ \ J. London Math. Soc. (3) 9 (1959), 373-387.
\item \ \ \ W.Raji, Modular Forms Representable As Eta Products, Dissertation Thesis (2006)
\item \ [7] G.V.Voskresenskaya, Finite Groups and Families of Modular Forms Associated
\item \ \ \ \ \ with Them, Mathematical Notes, 87, no.4(2010), 497–509
\item \ \ \ G.V.Voskresenskaya, Multiplicative Products of Dedekind Eta-Functions and Group
\item \ \ \ \ \ Representations, Mathematical Notes, vol.73, no.4(2003), 482–495.
\item [8] Y. Martin, Multiplicative Eta-Quotients. Trans.Amer.Math.Soc.348,
\item \ \ \ \ \ no.12 (1996), 4825-4856.
\item [9] H.R.Karadayi, M.Gungormez, Fundamental Weights, Permutation Weights and
\item \ \ \ \ \ Weyl Character Formula, J.Phys.A32:1701-1707(1999)
\item [10] M,Gungormez, H.R.Karadayi, Permutation Weights for $A_r^{(1)}$ Lie Algebras,
\item \ \ \ \ \  arXiv:1007.2718

\end